\documentstyle[12pt,epsf,psfig]{article}

\newcommand{\gsim}{\mathrel{\raisebox{-.6ex}{$\stackrel{\textstyle>}{\sim}$}}}
\def\beq{\begin{equation}}
\def\eeq{\end{equation}}
\def\bea{\begin{eqnarray}}
\def\eea{\end{eqnarray}}

\begin{document}

\thispagestyle{empty}

\font\fortssbx=cmssbx10 scaled \magstep2
\hbox to \hsize{
      \hfill$\vtop{
\hbox{MADPH-02-1273}
\hbox{UPR-0994T}}
$}

\vspace{.5in}

\begin{center}
{\large\bf No-go for detecting CP violation via\\
neutrinoless double beta decay}\\
\vskip 0.4cm
{V. Barger$^1$, S.L.~Glashow$^2$, P. Langacker$^3$ and D. Marfatia$^2$}
\\[.1cm]
$^1${\it Department of Physics, University of Wisconsin, Madison, WI
53706, USA}\\
$^2${\it Department of Physics, Boston University, Boston, MA 02215, USA}\\
$^3${\it Department of Physics and Astronomy, 
University of Pennsylvania,\\
Philadelphia, PA 19104, USA}\\
\end{center}

\vspace{.5in}

\begin{abstract}     

We present a necessary condition on the solar oscillation amplitude for 
CP violation to be detectable through neutrinoless double 
beta ($0\nu\beta\beta$) decay. It depends only on the fractional 
uncertainty in the
$\nu_e$--$\nu_e$ element of the neutrino mass matrix.
We demonstrate that even under very optimistic assumptions about
the sensitivity of future experiments to the absolute neutrino mass scale, 
and on the precision with which nuclear matrix elements that contribute to 
$0\nu\beta\beta$ decay are calculable, it will
be impossible to detect neutrino CP violation arising from Majorana phases.  

\end{abstract}

\thispagestyle{empty}
\newpage

If neutrinos are Majorana, the potentiality of detecting CP violation using
neutrinoless double beta decay exists~\cite{work}.
We consider this in a scenario wherein there are exactly three
left-handed neutrino states with Majorana masses. We derive a
necessary condition that involves the solar oscillation amplitude and the 
uncertainty in the $\nu_e$--$\nu_e$ element of the neutrino mass matrix 
that must be satisfied for CP violation to be detectable. Assuming that
this condition is satisfied and allowing for 
experimental and theoretical uncertainties that are unrealistically
small in some cases, we show that it will not be possible to detect neutrino 
CP violation through this process. 

The charged-current eigenstates are related to the
mass eigenstates by a unitary transformation
\beq
\left( \begin{array}{c} \nu_e \\ \nu_\mu \\ \nu_\tau \end{array} \right)
= U V \left( \begin{array}{c} \nu_1 \\ \nu_2 \\ \nu_3 \end{array} \right)
= \left( \begin{array}{ccc}
  c_{2} c_{3}                           & c_{2} s_{3}
& s_{2} e^{-i\delta} \\
- c_{1} s_{3} - s_{2} s_{1} c_{3} e^{i\delta} &   c_{1} c_{3} - s_{2} s_{1} s_{3} e^{i\delta}
& c_{2} s_{1} \\
  s_{1} s_{3} - s_{2} c_{1} c_{3} e^{i\delta} & - s_{1} c_{3} - s_{2} c_{1} s_{3} e^{i\delta}
& c_{2} c_{1} \\
\end{array} \right) V
\left( \begin{array}{c}
\nu_1 \\ \nu_2 \\ \nu_3
\end{array} \right) \,,
\label{eq:U}
\eeq
where $s_i$ and $c_i$ are the sines and cosines of $\theta_i$,
and $V$ is the diagonal matrix 
$(1,e^{i{\phi_2 \over 2}},e^{i({\phi_3\over 2}+\delta)}$).
In Eq.~(\ref{eq:U}), $\phi_2$ and $\phi_3$ are the Majorana phases
that are not measurable in neutrino oscillations and which
 are either $0$ or $\pi$ if CP is conserved.

The solar neutrino data allow two solutions at the 
3$\sigma$ C.L.~\cite{Barger:2002iv}: 
\bea
&&0.56 \le \sin^22\theta_3 \le 0.95\,, \ \ \ \ 2.0 \times
10^{-5} \le \Delta_s \le 2.3 \times 10^{-4}~{\rm{eV}}^2\,,\ \ \ \ ({\rm{LMA}})\\
&&0.80 \le \sin^22\theta_3 \le 0.94\,, \ \ \ \ 9.0 \times
10^{-8} \le \Delta_s \le 2.0 \times 10^{-7}~{\rm{eV}}^2\,,\ \ \ \ ({\rm{LOW}})
\eea
of which the LMA is the only solution at the 99\% C.L. No solution 
with $\theta_3\geq \pi/4$ \mbox{($\cos 2\theta_3\leq 0$)} is allowed at the 
5$\sigma$ C.L.~\cite{Barger:2002iv}.
Atmospheric neutrino data imply $\sin^22\theta_1 \ge 0.85$ and
$ 1.1 \times 10^{-3}$~eV$^2 \le \Delta_a \le 5 \times
10^{-3}$~eV$^2$ at the 99\% C.L.~\cite{Toshito:2001dk}. 
The CHOOZ reactor experiment imposes
the constraint $\sin^22\theta_{2} \le 0.1$ at the 95\% C.L.~\cite{CHOOZ}.
$\Delta_s$ and $\Delta_a$ are the 
mass-squared differences relevant to
solar and atmospheric neutrino oscillations, respectively.

vWe choose the mass ordering $m_1 < m_2 < m_3$ with $m_i$ non-negative. 
There are two possible
neutrino mass spectra:
\begin{eqnarray}
\Delta_s=m_2^2-m_1^2\,,&& \Delta_a=m_3^2-m_2^2\,,\ \ \ \ \ \ {\rm (normal\
  hierarchy),}\\
\Delta_s=m_3^2-m_2^2\,,&& \Delta_a=m_2^2-m_1^2\,,\ \ \ \ \ \ {\rm (inverted\
  hierarchy),}
\end{eqnarray}
where in either case $\Delta_a \gg \Delta_s$ in accord with experimental data.
 For the {\it normal hierarchy} (Case~I), mixing 
is given by
Eq.~(\ref{eq:U}).  
In this case solar neutrinos oscillate  primarily between the
two lighter mass eigenstates. 
  For the  {\it inverted hierarchy} (Case~II),  solar neutrinos
oscillate primarily between the two nearly degenerate 
heavier states. The oscillation parameters  
retain the same import as in Case~I if the columns
of $UV$ are permuted so that the third column takes the place of
the first column. 

It is well known that there are four combinations of the intrinsic 
CP parities of the mass eigenstates for which CP invariance holds~\cite{wolf}. 
If $\theta_2=0$, the number
of such combinations is reduced to two.
To minimize the  
confusion between CP violation and CP conservation, we
neglect $\theta_2$ which is already constrained to be small.  
This is justified as long as the mass spectrum is not hierarchical
($m_1 \ll m_2 \ll m_3$). Since experiments
will not be sensitive to the $M_{ee}$ ($< 0.01$ eV) expected for a 
hierarchical spectrum our assumption that $\theta_2$ is small is not only
sound but also optimistic from the viewpoint of detecting CP violation via
 $0\nu\beta\beta$ decay. Of course, if $\theta_2$ is small, long baseline
neutrino experiments will find it harder to detect CP violation resulting 
from $\delta$.
In this sense $0\nu\beta\beta$ decay and long baseline experiments are
complementary for detecting CP violation; if $\theta_2$ is close to the 
CHOOZ bound, long baseline experiments have a greater chance of detecting
CP violation and if $\theta_2$ is tiny, $0\nu\beta\beta$ decay experiments
have the better chance. 

Under the assumption that $\theta_2$ is negligible,
the $\nu_e$--$\nu_e$ element of the neutrino mass matrix~\cite{0nubb},
is
\bea
M_{ee} &=& | c^2 m_1 + s^2 m_2 e^{i\phi}| \,, \ \ \ {\rm~(Case~I)} \,,
\label{Mee1}\\
&=& |c^2 m_2 + s^2 m_3 e^{i\phi}| \,, \ \ \ {\rm~(Case~II)} \,,
\label{Mee2}
\eea
where we have dropped the subscripts on the sines and cosines of $\theta_3$ 
and on $\phi_2$, since only one mixing angle and one Majorana phase is
involved.
The masses $m_i$ may be determined from the lightest mass $m_1$ and the
mass--squared differences.
Since the solar mass--squared difference is
very small it can be ignored; then setting $m_1=m$ and $\Delta_a=\Delta$,
\bea
&& m_e = m_2 = m \,, \qquad m_3 = \sqrt{m^2 + \Delta} \,,
\ \ \  {\rm~(Case~I)} \,,
\label{eq:mass1}\\
&& m_e = m_2 = m_3  =  \sqrt{m^2 + \Delta} \,,
\ \ \ \ \ \ \ \ \ \ \ \ \ \,{\rm~(Case~II)} \,.
\label{eq:mass2}
\eea
Here, the mass of the electron neutrino $m_e$ is what tritium beta decay
experiments seek to measure. It 
is related to the sum of neutrino masses ($\Sigma = \Sigma m_i$) 
obtainable from data on weak lensing of galaxies and the CMB
via
\beq
\Sigma = 2 m_e + \sqrt{m_e^2 \pm \Delta} \,,
\label{eq:sum}
\eeq
where the plus sign applies to the normal hierarchy and the minus sign
to the inverted hierarchy. Using Eqs.~(\ref{Mee1}-\ref{eq:mass2}), the
Majorana phase in either case is given by
\beq
\cos \phi=1-{2\over \sin^2 2\theta} \bigg(1-{M_{ee}^2 \over m_e^2}\bigg) \,.
\eeq
The second term on the right hand side quantifies the deviation of $\phi$
from 0. If $M_{ee}/m_e=1$, $\phi=0$ and if $M_{ee}/m_e=\cos 2\theta$, 
$\phi=\pi$. Thus,
\beq
\cos 2\theta \leq {M_{ee}\over m_e} \leq 1   \,,
\label{limits}
\eeq
with the boundaries of the interval corresponding to CP conservation. 
A measurement of $M_{ee}/m_e$ that excludes the
boundaries constitutes a detection of CP violation. 
A larger solar amplitude (a wider interval) is
therefore more conducive to such a measurement. Let us evaluate the minimum
$\sin^2 2\theta$ for which CP violation can be detected assuming
a measurement $M_{ee}(1\pm x)$, where 
$x$ is obtained by summing the 
theoretical uncertainty in the $0\nu\beta\beta$ nuclear matrix elements and
the experimental uncertainty
in quadrature. 
Then, for CP violation to be detectable the necessary condition is
\beq
{\cos 2\theta \over 1-x}< {M_{ee}\over m_e} < {1\over 1+x}    \,,
\eeq
or
\beq
\sin^2 2\theta > 1- \bigg({1-x \over 1+x}\bigg)^2\,.
\label{condition}
\eeq
The current solar data require $\sin^2 2\theta$ to be smaller than 0.95 at the
3$\sigma$ C.L. Thus, $x$ must be smaller than 0.63. It is a difficult 
task to reduce the factor of $3$ uncertainty 
 on the nuclear matrix elements (corresponding to $M_{ee}(1^{+2.0}_{-0.7})$) 
to such a degree 
especially since a reliable method for estimating the uncertainty 
does not exist{\footnote{The often quoted 
factor of 3 uncertainty represents the range 
of calculated values of the matrix elements available in the 
literature.}}~\cite{vogel}. 
Conversely, for a realistically achievable improvement in $x$, 
it is unlikely
that the solar amplitude is sufficiently large so as to satisfy
  Eq.~(\ref{condition}).

In what follows, we consider the remote possibility that 
Eq.~(\ref{condition}) is in fact satisfied, and show that it is not 
sufficient to detect CP violation.

We work under the following assumptions about the experimental and
theoretical developments that might occur by 2020:

\begin{enumerate}
\item{Experiments like GENIUS~\cite{genius} and EXO~\cite{exo}
 are sensitive to $M_{ee}$ above 0.01 eV with a 25\% experimental 
uncertainty~\cite{vogel} 
and are therefore not sensitive to the hierarchical neutrino mass spectrum.
Also, since $M_{ee}\leq m_e$, we can draw meaningful 
conclusions only for $m_e\gsim 0.01$ eV or equivalently 
$\Sigma \gsim 0.08$ eV.}
\item{A breakthrough in the evaluation of the  
nuclear matrix elements has allowed an estimate of their uncertainty. 
The factor of $3$ uncertainty that has plagued the
 matrix elements is reduced so that the combined theoretical and 
experimental uncertainty on $M_{ee}$ is $x=0.5$. Then Eq.~(\ref{condition}) is satisfied
for $\sin^2 2\theta> 0.89$.}
\item{Tritium beta decay experiments like KATRIN are sensitive to 
$m_e$ above 0.35 eV with an uncertainty of 
0.08 eV$^2$ on $m_e^2$~\cite{KATRIN}. }
\item{Weak lensing of galaxies by large scale structure in conjunction with 
CMB data can measure $\Sigma$ to an uncertainty of 0.04 eV~\cite{hu}.  }
\item{The KamLAND (Borexino) experiment has determined the solar oscillation
amplitude to be $0.95\pm 0.04$ where the precision in the 
LMA (LOW) region was estimated in 
Ref.~\cite{barger} (\cite{deGouvea:1999xe}).} 
\item{The JHF-Kamioka neutrino project has measured $\Delta$ 
to be $3\times 10^{-3}$ eV$^2$ to within 3\% and has constrained 
$\sin^2 \theta_2$ to be smaller than $2\times 10^{-3}$~\cite{Itow:2001ee}.}
\item{The neutrino mass hierarchy is determined either from a superbeam
experiment or from supernova neutrinos.} 
\end{enumerate}

While some of these assumptions are overly optimistic, they serve well
to show once and for all that it is not possible to detect CP violation from
 $0\nu\beta\beta$ decay in the foreseeable future.

Our assumptions clearly suggest that cosmology will not only probe smaller 
neutrino masses, but also with greater precision than tritium beta decay 
experiments. However, to have the possibility of an 
 independent confirmation from a 
table-top experiment (in the regime of common sensitivity) 
is invaluable. 

In our quantitative analysis, we assume the precision on 
$\Sigma$ expected from cosmology. 
We fix $\Delta=3\times 10^{-3}$ eV$^2$
and assign no uncertainty to its value. We suppose that precise measurements 
of $\Sigma$ and $M_{ee}$ will be made such that the 
central value of $M_{ee}/m_e$ lies in the interval of Eq.~(\ref{limits}).
One expects the extent to which CP violation can be detected to be 
dependent upon how
close the central value of $M_{ee}/m_e$ is to the middle of the interval.

\begin{figure}[ht]
\centering\leavevmode
\psfig{file=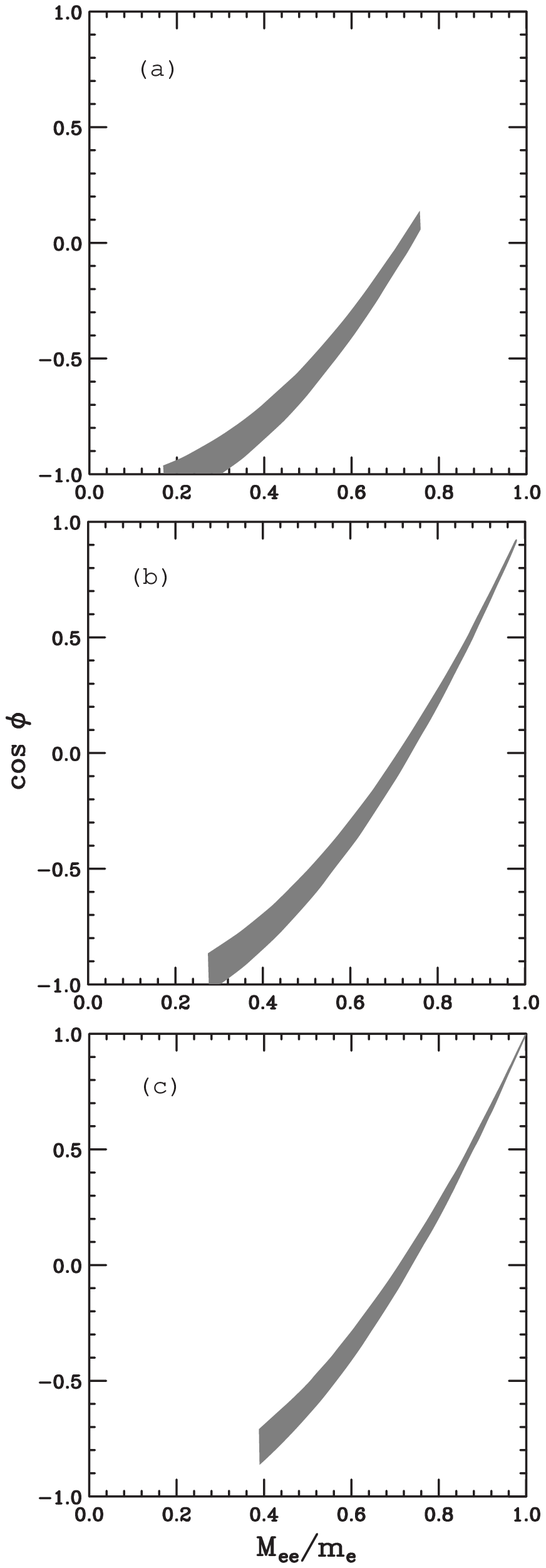,width=7cm,height=17cm}\\
\medskip
\caption{``$1\sigma$'' bands in the $M_{ee}/m_e$--$\cos \phi$ plane for three 
possible measurements assuming the normal hierarchy: 
(a) $\Sigma=0.24 \pm 0.04$ eV, $M_{ee}=0.03(1 \pm 0.5)$ eV with
the central value of $M_{ee}/m_e=0.4$, 
(b) $\Sigma=0.51 \pm 0.04$ eV, $M_{ee}=0.10(1 \pm 0.5)$ eV with
the central value of $M_{ee}/m_e=0.6$,
(c) $\Sigma=1.12 \pm 0.04$ eV, $M_{ee}=0.30(1 \pm 0.5)$ eV with
the central value of $M_{ee}/m_e=0.8$.
Results are shown for the best possible solar amplitude 
(allowed by present solar data at 3$\sigma$) 
for the detection of CP violation, 
$\sin^2 2\theta=0.95 \pm 0.04$. $\Delta$ is fixed at $3\times 10^{-3}$ eV$^2$.}
\label{fig1}
\end{figure}

For hypothetical measurements of $\Sigma$ and $M_{ee}$, the
regions allowed by scans within the $1\sigma$ uncertainties assumed for each
of the measurements are shown in Fig.~\ref{fig1} for the normal hierarchy.
We have chosen the central values of 
$M_{ee}/m_e$ to be 0.4, 0.6 and 0.8 so that they are not too close to
either $\cos 2\theta$ ($=0.22$) or $1$ for which  $\cos \phi=-1$ 
and $\cos \phi=1$,
respectively, are unavoidable. 
For $\Sigma \gg \sqrt{\Delta}$, the regions for both hierarchies
are almost identical. For the case in which $\Sigma=0.24$ eV, the
regions for the inverted hierarchy do not extend to as high values of 
$M_{ee}/m_e$ as for the normal hierarchy, but qualitatively they are 
the same in that $\cos \phi=-1$ is allowed. 

We emphasize that while CP violation
is not detectable via $0\nu\beta\beta$ decay, if the solar amplitude is 
found to be larger than current 
solar data suggest and if the precision on the various measurements 
 and the refinement of the calculation of the nuclear matrix elements
assumed by us is achieved, it might be possible to determine if $\phi$
is closer to $0$ or to $\pi$ (see Fig.~\ref{fig1}).

\vskip 0.4in
\noindent
{\it Acknowledgements.}  
We thank S. Dodelson for drawing our attention to Ref.~\cite{hu}. 
This work was supported in part by the NSF under
grant No.~NSF-PHY-0099529, in part by the
U.S. Department of Energy under grant
Nos.~DE-FG02-91ER40676, 
~DE-FG02-95ER40896 and~DOE-EY-76-02-3071, and in part by the
Wisconsin Alumni Research Foundation.

\clearpage

\end{document}